\documentclass[12pt,preprint]{aastex}
\usepackage{graphicx}
\usepackage{longtable}
\usepackage[CaptionAfterwards]{fltpage}

\shorttitle{HD 97658b transits}

\shortauthors{Dragomir et al.}

\begin{document}

\renewcommand{\thefootnote}{\fnsymbol{footnote}}

\title{MOST\footnote{Based on data from the MOST satellite, a Canadian Space Agency mission operated by Microsatellite Systems Canada Inc. (MSCI; former Dynacon Inc.) and the Universities of Toronto and British Columbia, with the assistance of the University of Vienna.}  detects transits of HD 97658\lowercase{b}, a warm, likely volatile-rich super-Earth}

\author{
  Diana Dragomir\altaffilmark{1,2},
  Jaymie M. Matthews\altaffilmark{3},
  Jason D. Eastman\altaffilmark{1,2},
  Chris Cameron\altaffilmark{4,5},
  Andrew W. Howard\altaffilmark{6},
  David B. Guenther\altaffilmark{7},
   Rainer Kuschnig\altaffilmark{8},
  Anthony F. J. Moffat\altaffilmark{9},
  Jason F. Rowe\altaffilmark{10,11},
  Slavek M. Rucinski\altaffilmark{12},
  Dimitar Sasselov\altaffilmark{13}, 
  Werner W. Weiss\altaffilmark{8}
}

\email{diana@lcogt.net}
\altaffiltext{1}{Las Cumbres Observatory Global Telescope Network, 6740 Cortona Dr. suite 102, Goleta, CA 93117, USA}
\altaffiltext{2}{ Department of Physics, Broida Hall, UC Santa Barbara, CA}
\altaffiltext{3}{Department of Physics and Astronomy, University of British Columbia, Vancouver, BC V6T1Z1, Canada}
\altaffiltext{4}{Department of Mathematics, Physics \& Geology, Cape Breton University, 1250 Grand Lake Road, Sydney, Nova Scotia, Canada, B1P 6L2}
\altaffiltext{5}{43 Judy Anne Crt., Lower Sackville, Nova Scotia, Canada, B4C 3X8}
\altaffiltext{6}{Institute for Astronomy, University of Hawaii, 2680 Woodlawn Drive, Honolulu, HI 96822, USA}
\altaffiltext{7}{Department of Astronomy and Physics, St. Mary's University Halifax, NS B3H 3C3, Canada}
\altaffiltext{8}{Universit\"{a}t Wien, Institut f\"{u}r Astronomie, T\"{u}rkenschanzstrasse 17, AÐ1180 Wien, Austria}
\altaffiltext{9}{D\'{e}pt de physique, Univ de Montr\'{e}al C.P. 6128, Succ. Centre-Ville, Montr\'{e}al, QC H3C 3J7, and Obs. du mont M\'{e}gantic, Canada} 
\altaffiltext{10}{SETI Institute, 189 Bernardo Ave, Mountain View, CA 94043, USA}
\altaffiltext{11}{NASA Ames Research Center, Moffett Field, CA 94035}
\altaffiltext{12}{Department of Astronomy and Astrophysics, University of Toronto, 50 St. George Street, Toronto, ON M5S 3H4, Canada}
\altaffiltext{13}{Harvard-Smithsonian Center for Astrophysics, 60 Garden Street, Cambridge, MA 02138, USA}

\begin{abstract}

Through photometric monitoring of the extended transit window of HD 97658b with the MOST space telescope, we have found that this exoplanet transits with an ephemeris consistent with that predicted from radial velocity measurements. The mid-transit times are $5.6\sigma$ earlier than those of the unverified transit-like signals reported in 2011, and we find no connection between the two sets of events. The transit depth together with our determined stellar radius ($R_\star = 0.703^{+0.039}_{-0.034} R_\odot$) indicates a 2.34$^{+0.18}_{-0.15}$ $R_\earth$ super-Earth. When combined with the radial velocity determined mass of 7.86 $\pm 0.73$ $M_\earth$, our radius measure allows us to derive a planet density of 3.44$^{+0.91}_{-0.82}$ g cm$^{-3}$. Models suggest that a planet with our measured density has a rocky core that is enveloped in an atmosphere composed of lighter elements. The star of the HD 97658 system is the second brightest known to host a transiting super-Earth, facilitating follow-up studies of this not easily daunted, warm and likely volatile-rich exoplanet. 

\end{abstract}


\section{Introduction}

Transiting super-Earth exoplanets are an important and interesting class of planets to study for two main reasons: no super-Earths exist in the Solar System, and their masses and radii generally allow for a significant range of compositions. The most common way to home in on the composition of a super-Earth is by precisely determining its mass and radius. The {\it Kepler} mission has been extremely successful in finding super-Earths and measuring their radii with unprecedented precision. However, the majority of the stars hosting these planets are too faint to allow for the precise radial velocity (RV) measurements that most effectively determine the mass of an exoplanet. Spectroscopic observations of these exoplanets' atmospheres also require bright host stars. For these reasons, super-Earths transiting bright stars like HD 97658 are essential for the characterization of this class of exoplanet.

The planet orbiting the K1 dwarf HD 97658 was announced by \cite{How11} with a minimum mass of 8.2 $\pm$ 1.2 $M_\earth$ and an orbital period of 9.494 $\pm$ 0.005 days. This potential super-Earth has already been photometrically searched for transits since its discovery. Transits announced in 2011 \citep{Hen11} were later shown to be spurious \citep{Dragomir97} using high-precision MOST (\citealt{Mat04}, \citealt{Wal03}) photometry. G. Henry was also unable to confirm the transits with additional Automated Photometric Telescope (APT) photometry acquired during the 2012 observing season (private communication).
 
 \setcounter{footnote}{0}
 \renewcommand{\thefootnote}{\arabic{footnote}} 
 
The MOST photometry that was used to reject those events only covered the RV transit window between +0.55 and +3.6$\sigma$ of the predicted mid-transit time\footnote{The transit window is the time span during which a transit is predicted to occur, calculated from the uncertainties on the orbital period and those on the predicted mid-transit time.} (i.e., 71\% of the mid-transit time's posterior probability). We completed the coverage of the 3$\sigma$ transit window by scheduling another set of MOST observations in April 2012, covering -3.7 to +1.5$\sigma$ of the predicted RV transit window (i.e., 99.97\% of the mid-transit time's posterior probability, when combined with the previous results). We noticed an intriguing dip in this light curve, but were unable to follow it up because the star had left the satellite's Continuous Viewing Zone (CVZ). We were able to confirm that the candidate signal is real by re-visiting the system in 2013 and observing the signal at the expected time during four additional consecutive transit windows.

In this Letter we announce the discovery of HD 97658b transits, the depth of which indicates, together with the mass obtained from the RVs, that the planet is a super-Earth. We describe our data reduction procedure, our analysis of the photometry and our conclusions in the sections that follow.

\section{Observations}

For consistency, all our times for both the RV and photometric data sets are in BJD$_{TDB}$ \citep{East10}. Below we describe the two sets of observations.

\subsection{Keck Radial Velocity Measurements}

Since the publication of \cite{Dragomir97}, we have obtained four new Keck HIRES RV measurements. These were acquired and reduced using the same techniques as in \cite{How11}. We combined the new measurements with the existing radial velocities and excluded the same three outliers as in \cite{Dragomir97}. In total, we used 171 radial velocities for the analysis described in Section 3. The full set of RVs are listed in Table 2.

\subsection{MOST Photometry}

In an effort to monitor as much of the radial-velocity predicted transit window as possible and wrap up the search for transits of HD 97658b, we have acquired MOST observations of the system in addition to those published in \cite{Dragomir97}. The first of these new data were acquired on April 11-12, 2012 and cover the RV transit window between approximately -3.7 and +1.5$\sigma$. This transit window was computed from the ephemeris reported in \cite{Dragomir97}. A shallow dip can be seen at BJD of about 2456029.7 or approximately 1.1$\sigma$ before the predicted mid-transit time. It was not possible to obtain further MOST photometry of the system in order to verify the repeatibility of this candidate before it left the satellite's CVZ. As soon as HD97658 re-entered the MOST CVZ in 2013 and we were able to interrupt primary target observations, we re-observed it during four transit windows based on the mid-transit time of the 2012 candidate. Those data were acquired on March 10, 19, 29 and April 7, 2013. The exposure times were 1.5 s, and the observations were stacked on board the satellite in groups of 21 for a total integration time of 32 s per data point.

The light curves covering each transit window were reduced individually. The raw photometry was extracted using aperture photometry \citep{Row08}. Outliers more than 3$\sigma$ from the mean of each light curve were clipped. The resulting magnitudes were then de-correlated from the sky background using 4th or 5th order polynomials, and from x and y position on the CCD using 2nd or 3rd order polynomials. After these steps, a straylight variation at the orbital period of the satellite remains. This variation is filtered by folding each light curve on this 101.4-minute period, computing a running average from this phased photometry, and removing the resulting waveform from the corresponding light curve. 

The five light curves are shown in Figure \ref{fig:transits1}. The 2012 observations and the last set of 2013 observations were acquired when the star was on the edge of or slightly outside the CVZ. Therefore, a star in the CVZ had to be used as a switch target during part of every MOST orbit, leading to the gaps that are visible in each of those two light curves. The increasing flux portions at the beginning of the three middle light curves correlate with a sudden change in temperature of the pre-amp board. This occurs when the satellite switches between two targets that are far apart from each other on the sky. 

\begin{FPfigure}
\centering
\includegraphics[scale=0.071]{HD97658_individual.pdf}
\caption{MOST light curves of HD 97658 acquired, from top to bottom, on April 11-12, 2012, March 10, 19, 29 and April 7, 2013. The tire-track pattern in the first and last light curves is due to the satellite alternating between HD 97658b and another target. The red vertical bars correspond to the best-fit mid-transit time obtained from fitting the three continuous transits. The grey solid vertical bars represent the mid-transit predicted from RVs only, and the dotted grey vertical bars enclose the RV-only 1$\sigma$ transit window. The times where transits predicted by \cite{Hen11} would occur, if they were real, are indicated in each time series by a black arrow. These times were obtained by propagating forward the mid-transit time reported in \cite{Hen11} {\it using our more precise estimate of the planet's orbital period} from Table 1. The increase in flux at the beginning of the three middle light curves is due to a change in temperature which occurs when the satellite switches between two targets that are far apart from each other. }
\label{fig:transits1}
\end{FPfigure}
\clearpage

\section{Analysis}

Every one of the five light curves shows a transit-like event, spaced by $\approx$9.49 d and occurring about 1.2 - 1.3$\sigma$ earlier than the radial-velocity predicted mid-transit time (the solid grey vertical bars). The extent of the 1$\sigma$ RV-predicted transit window is shown (enclosed by pairs of dotted grey vertical bars), as well as the predicted time of the \cite{Hen11} events (if they were real), propagated to the epochs of our transits {\it using our more precise estimate of the orbital period} listed in Table 1. The first and last light curves suffered from increased scatter from straylight, and the instrument's pointing stability from one HD 97658 visit to the next (within a given light curve) is not optimized because of the alternating target setup. In addition, the last step of our reduction routine (the removal of straylight artifacts) is not as effective for interrupted light curves. Photometry during part of every MOST orbit is missing, and the re-constructed waveform is not as accurate as for continuous light curves. The effect also depends on the phase and fraction of the MOST orbit that is missing. This in turn affects the shape of shallow signals with durations on the order of one or a couple of satellite orbits. 

As an additional check of the transits' authenticity, we have inspected the light curves of the two other stars in MOST's field of view of HD 97658. We do not observe any brightness variations in those two stars resembling the HD 97658b transits in either duration or phase. 

Before fitting the data, we quantified the correlated noise present in the light curves by following the method described in \cite{Win11} with minor modifications. We binned the out-of-transit photometry using bin sizes between 5 and 60 minutes and compared the rms of each binned light curve to what we would expect it to be if the light curve only contained white noise. We multiplied the photometric uncertainties (determined during the aperture photometry extraction) by 2.5, the largest
value of the scaling factor found during these comparison tests.

We fit the data with EXOFAST \citep{East13}, a MCMC algorithm that can simultaneously model RVs and photometry. We used a modified version of the algorithm which employs Yonsei-Yale isochrones \citep{Yi, Dem} together with spectroscopically derived values for stellar effective temperature ($T_{eff}$) and metallicity ($[Fe/H]$), as well as the transit photometry to constrain the stellar parameters. We used $T_{eff}=5119 \pm 44 K$ and $[Fe/H]=-0.3 \pm 0.03$ from \cite{Hen11}. The algorithm uses the values and uncertainties of these two parameters to determine the stellar mass and radius via isochrone analysis, the uncertainties of which then propagate to the planetary mass and radius. Therefore it is important that the uncertainties on $T_{eff}$ and $[Fe/H]$ are not underestimated. When comparing values from different catalogs, there is evidence that uncertainties on individual metallicity estimates are often underestimated \citep{Hinkel}. Further, we found that the values of these two parameters differ by 1-2$\sigma$ between \cite{How11} and \cite{Hen11}. Therefore, we scaled these uncertainties upwards, to 50 K for the effective temperature and to 0.08 dex for the metallicity \citep{Buch}, for the EXOFAST fit. We used quadratic limb darkening coefficients for the MOST bandpass of $u_{1}=0.621 \pm 0.050$ and $u_{2}=0.141 \pm 0.050$ generated using the models of \cite{CK} (A. Prsa, private communication).

\begin{figure}[!h]
\begin{center}
\includegraphics[scale=0.18]{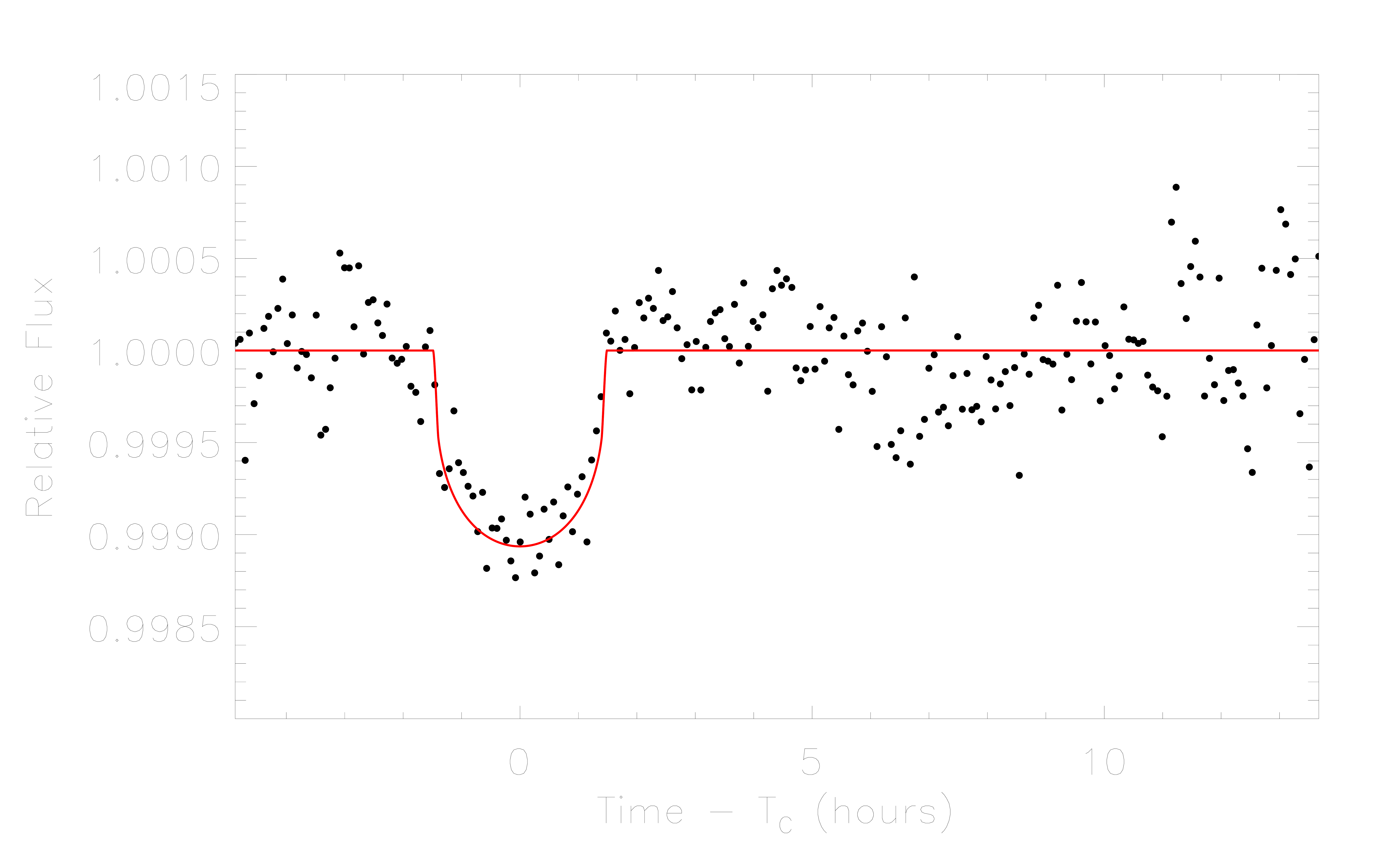}
\caption{\label{fig:transits2} MOST photometry of the three HD 97658b continuous transit light curves, folded on the best-fit (median) period from the EXOFAST fit (9.4909 days) and averaged in 5-min bins. The red curve is the best-fitting transit model based on the EXOFAST fit of the three continuous transits.}
\end{center}
\end{figure}

We fit the photometry together with the RVs to ensure the two data sets were consistent with each other. We carried out one run using all five transits, and another using only the three continuous transits. The former run resulted in a planetary radius just over 1$\sigma$ shallower than the latter run. For the reasons discussed at the beginning of this section, we chose to use the results from the run based only on the continuous transits for the remainder of this Letter. 

It has been shown (\citealt{Win11}, \citealt{Dragomir97}) that the MOST reduction pipeline can sometimes suppress the depth of a transit signal. We have carried out a transit injection and recovery test to quantify this effect for the HD 97658 time series. Simulated limb-darkened transits corresponding to a planet with radius 2.23 $R_\earth$ (the value obtained from the EXOFAST fit) were inserted in the raw photometry at 100 randomly distributed orbital phases overlapping with out-of-transit sections of the three continuous light curves. The photometry containing the simulated transits was then reduced following the same steps as for the unmodified photometry. The depths of the recovered transits were on average suppressed by 10\%, corresponding to a 5\% suppression in the planetary radius. We increased the planetary radius output by EXOFAST by this percentage, and adjusted the planetary density and surface gravity accordingly.

The final values for the stellar and planetary parameters of this fit are listed in Table 1. The folded transit based on the three continuous light curves, binned in 5-min bins, is shown in Figure \ref{fig:transits2}. To produce the phased time series, we omitted the first 0.1 days of each light curve which were affected by the temperature change described at the end of Section 2.2. 

We note that we find the uncertainties on the planetary radius are only 13\% larger than if the unscaled photometric uncertainties were used. This indicates that the uncertainty in the stellar radius is the dominating factor on the precision to which we can measure the planet's size with the current data. The stellar mass and radius obtained from the EXOFAST fit are in excellent agreement with those quoted in \cite{Hen11}. Our value of $a/R_\star$ is consistent with the value of this ratio derived using $a$ from the RVs alone and $R_\star$. Finally, we compare the value of log$g$ determined by EXOFAST from the photometry and stellar parameters (4.618$^{+0.036}_{-0.041}$) with the spectroscopic log$g$ from \cite{Hen11} (4.52 $\pm$ 0.06). \cite{Buch} show that the noise floor for spectroscopic log$g$ dominated by uncertainties in stellar models is 0.1. Using this value as the uncertainty on the spectroscopic log$g$ for this star, we find that the photometric and spectroscopic log$g$ values agree to within 1$\sigma$.

\begin{figure}[!h]
\begin{center}
\includegraphics[scale=0.6]{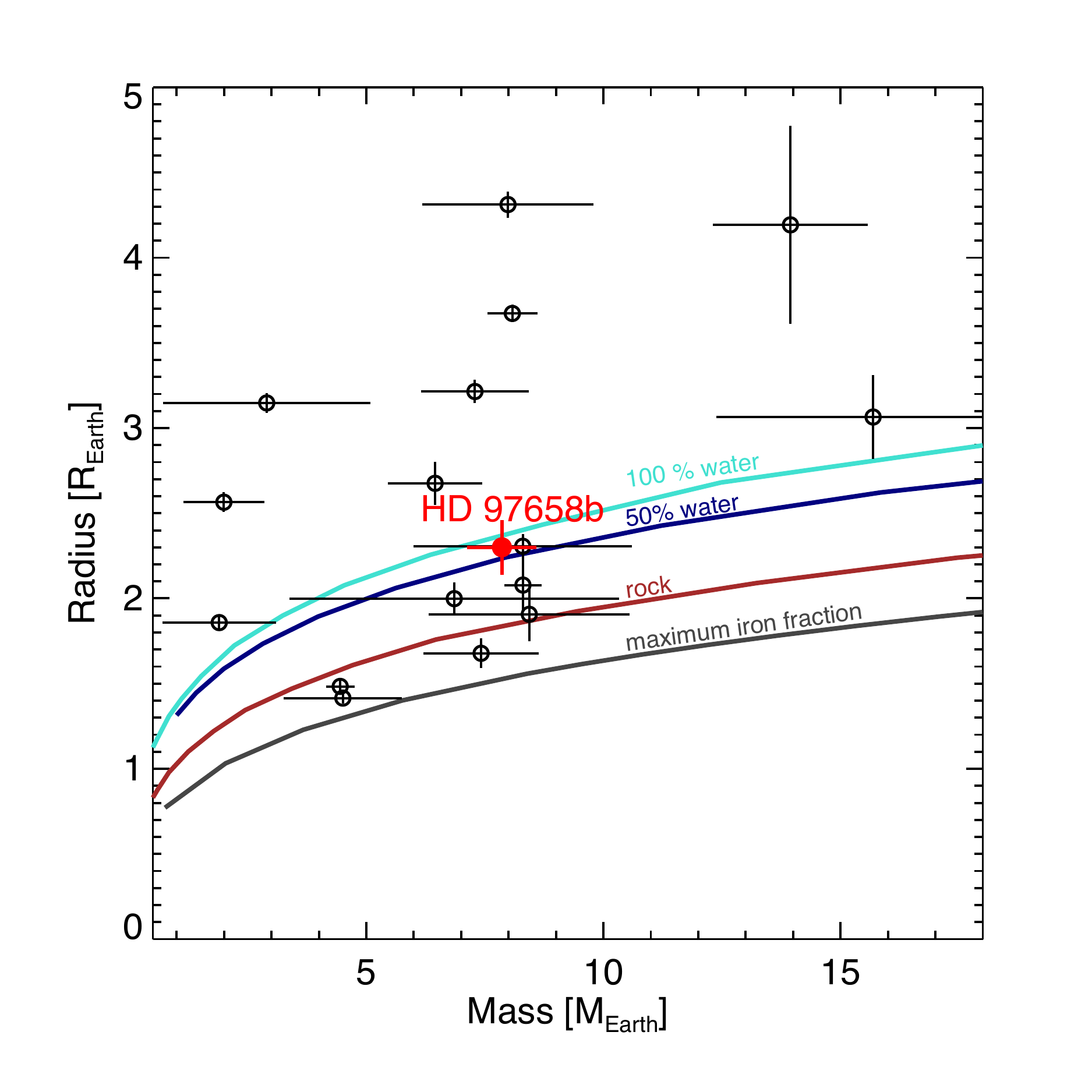}
\caption{Mass-radius diagram for currently known transiting super-Earths with masses measured either by RVs or TTVs. Planetary parameters were obtained from the Exoplanet Orbit Database at exoplanets.org \citep{Wri11}. Density model curves are shown for 100\% water, 50\% water/40\% silicate mantle/6\% iron core, and rock (silicate) planets \citep{Seager3}. The maximum iron fraction curve corresponds to planets with minimum radius defined by the maximum mantle stripping limit \citep{Marcus}.)
planets.}
\label{fig:mr}
\end{center}
\end{figure}

\begin{figure}[!h]
\begin{center}
\includegraphics[scale=0.6]{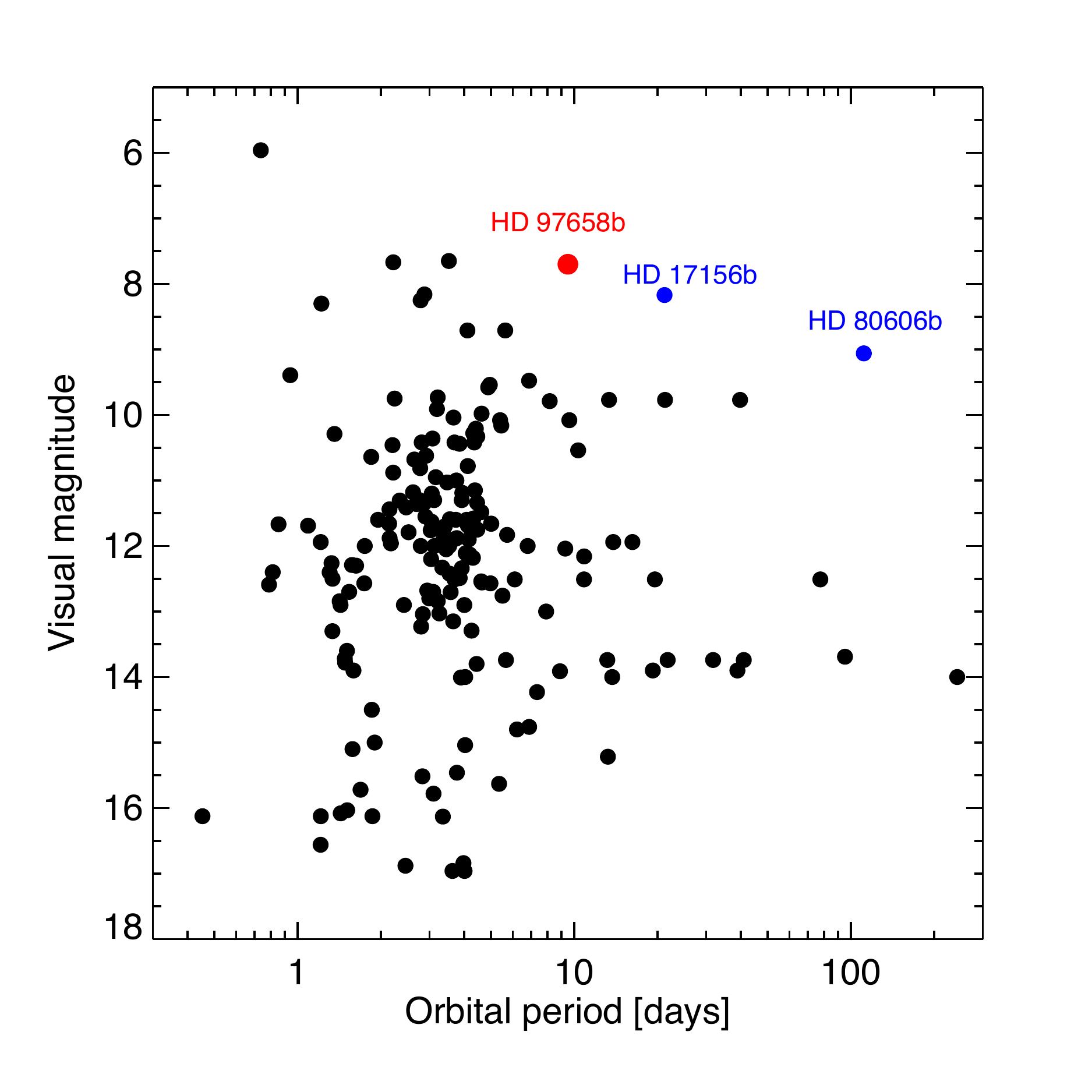}
\caption{The orbital periods of known transiting planets as a function of their host star V magnitude. Planets with intermediate or long periods ($\gtrapprox$ 6 days) orbiting stars brighter than V = 9 are shown in blue. HD 97658b (in red) now joins their ranks.}
\label{fig:magp}
\end{center}
\end{figure}

\section{Discussion}

We have carried out a search for transits of HD 97658b throughout its 3$\sigma$ RV-predicted transit window. We have discovered that the planet does cross the disk of its host star, allowing us to measure its size and therefore its density. The transits we have detected occur approximately 6$\sigma$ earlier than the transit-like signals reported in \cite{Hen11}. Propagating our mid-transit time backward to spring of 2011 (the epoch of the \citealt{Hen11} observations) indicates that transits are predicted to have occurred 16 $\pm$ 3 hours earlier than the transit-like signals observed by \cite{Hen11}, so our 3$\sigma$ transit window does not overlap with theirs. Further, our derived planetary radius is $>$3$\sigma$ smaller. For these reasons, we conclude that the transits announced in this paper bear no connection to the previously announced transit-like signals. 

HD 97658b has a radius of $2.34^{+0.18}_{-0.15} R_{\oplus}$, slightly larger than that of 55 Cnc e \citep{Win11}. Figure \ref{fig:mr} shows the mass and radius of HD 97658b relative to those of other known transiting super-Earths. Its density of $3.44^{+0.91}_{-0.82}$ g cm$^{-3}$ suggests the planet is probably not solely rocky. If it is composed of a rocky core, this core is most likely surrounded by an atmosphere of volatiles, by which we mean planetary ingredients lighter than just rock and iron. The mass and radius of HD 97658b are very similar to those of Kepler-68b, a planet in a multiple system with a period of 5.4 days \citep{Gilli}. Of the two, HD 97658b is significantly less irradiated, a characteristic which supports the existence of light elements such as hydrogen or helium in its atmosphere. Indeed, its zero-albedo equilibrium temperature is $\sim$1030 K assuming no heat redistribution, and $\sim$ 730 K for even heat redistribution. However, the measured density of this super-Earth is also consistent with a water planet.

HD 97658b is the second super-Earth known to transit a very bright ($V = 7.7$) star. Figure \ref{fig:magp} shows the orbital period of known transiting planets as a function of the magnitude of their host star. Of the now three exoplanets in the sparsely populated upper right area of the diagram, HD 97658b is the only super-Earth. It is enlightening to study how the structure and composition of warm super-Earths differs from those of their hotter counterparts. The brightness of HD 97658 makes this exoplanet system ideal for such investigations. We encourage follow-up observations of this system to more precisely constrain the planet's physical parameters and to begin probing its atmosphere. In fact, HD 97658b is an ideal candidate for atmospheric characterization with the James Webb Space Telescope (JWST; \citealt{Seager4, Shab11}).

More exoplanets will be found to transit bright stars through systematic photometric monitoring of known RV planets by projects such as the MOST and Spitzer super-Earth transit searches. The Transit Ephemeris Refinement and Monitoring Survey (TERMS; \citealt{Kane09, Drag12}) in particular will help further populate the upper right corner of Figure \ref{fig:magp} by searching for transits of RV planets with intermediate and long periods.

We have learned from the Kepler mission that super-Earths exist in multiple planet systems. We believe it is worthwhile to continue the radial velocity monitoring of the HD 97658 system in order to search for additional planetary companions. 

To close, the 4\% {\it a priori} transit probability of HD 97658b reminds us of the impartial nature of statistics: all probabilities, no matter how small, count in the race toward 100 percent.

\section{Acknowledgments}

We are grateful to Michael Gillon, Kaspar von Braun, Dan Fabrycky, Darin Ragozzine, Jean Schneider and Josh Winn for useful suggestions and assistance with Figures 2 and 3. We thank Peter McCullough, Heather Knutson and especially the anonymous referee for feedback which has helped improve this manuscript. We also thank Geoff Marcy, Debra Fischer, John Johnson, Jason Wright, Howard Isaacson and other Keck-HIRES observers from the California Planet Search. Finally, the authors wish to extend special thanks to those of Hawai`ian ancestry on whose sacred mountain of Mauna Kea we are privileged to be guests.  Without their generous hospitality, the Keck observations presented herein would not have been possible.

The Natural Sciences and Engineering Research Council of Canada supports the research of DBG, JMM, AFJM and SMR. Additional support for AFJM comes from FQRNT (Qu\'ebec). RK and WWW were supported by the Austrian Science Fund (P22691-N16) and by the Austrian Research Promotion Agency-ALR. 



\newcommand{\bjdtdb}{\ensuremath{\rm {BJD_{TDB}}}}
\newcommand{\feh}{\ensuremath{\left[{\rm {\it Fe}}/{\rm {\it H}}\right]}}
\newcommand{\teff}{\ensuremath{{\it T}_{\rm {\it eff}}}}
\newcommand{\ecosw}{\ensuremath{e\cos{\omega_*}}}
\newcommand{\esinw}{\ensuremath{e\sin{\omega_*}}}
\newcommand{\msun}{\ensuremath{\,M_\Sun}}
\newcommand{\rsun}{\ensuremath{\,R_\Sun}}
\newcommand{\lsun}{\ensuremath{\,L_\Sun}}
\newcommand{\mj}{\ensuremath{\,M_{\rm J}}}
\newcommand{\rj}{\ensuremath{\,R_{\rm J}}}
\newcommand{\me}{\ensuremath{\,M_{\rm E}}}
\newcommand{\re}{\ensuremath{\,R_{\rm E}}}
\newcommand{\fave}{\langle F \rangle}
\newcommand{\fluxcgs}{10$^9$ erg s$^{-1}$ cm$^{-2}$}
\begin{deluxetable}{lcc}
\tabletypesize{\footnotesize}
\tablecaption{Median values and 68\% confidence interval for the HD97658 system}
\tablewidth{0pt}
\tablehead{\colhead{~~~Parameter} & \colhead{Units} & \colhead{Value}}
\startdata
\sidehead{Stellar Parameters:}
	~~~${\it V}$ {\it mag} \dotfill &{\it Apparent V magnitude}\dotfill & ${\it 7.7}$\\
		~~~${\it u_1}$\dotfill &{\it linear limb-darkening coeff}\dotfill & ${\it 0.621}\pm{\it 0.050}$\\
           ~~~${\it u_2}$\dotfill &{\it quadratic limb-darkening coeff}\dotfill & ${\it 0.141}\pm{\it 0.050}$\\
                    ~~~$\teff$\dotfill &{\it Effective temperature (K)}\dotfill & ${\it 5119}\pm{\it 50}$\\
                              ~~~$\feh$\dotfill &{\it Metallicity}\dotfill & ${\it-0.30}_{{\it-0.08}}^{{\it+0.08}}$\\
                           ~~~$M_{*}$\dotfill &Mass (\msun)\dotfill & $0.747_{-0.030}^{+0.031}$\\
                         ~~~$R_{*}$\dotfill &Radius (\rsun)\dotfill & $0.703_{-0.030}^{+0.035}$\\
                         ~~~$\rho_*$\dotfill &Density (cgs)\dotfill & $3.04_{-0.39}^{+0.38}$\\
              ~~~$\log(g_*)$\dotfill &Surface gravity (cgs)\dotfill & $4.618_{-0.039}^{+0.034}$\\
     
\sidehead{Planetary Parameters:}
                               ~~~$e$\dotfill &Eccentricity\dotfill & $0.063_{-0.044}^{+0.059}$\\
    ~~~$\omega_*$\dotfill &Argument of periastron (degrees)\dotfill & $-9_{-63}^{+67}$\\
                              ~~~$P$\dotfill &Period (days)\dotfill & $9.4909_{-0.0015}^{+0.0016}$\\
                       ~~~$a$\dotfill &Semi-major axis (AU)\dotfill & $0.0796\pm0.0011$\\
                             ~~~$M_{P}$\dotfill &Mass ($M_{\oplus}$)\dotfill & $7.86\pm0.73$\\
                           ~~~$R_{P}$\dotfill &Radius ($R_{\oplus}$)\dotfill & $2.341_{-0.15}^{+0.17}$\\
                       ~~~$\rho_{P}$\dotfill &Density (cgs)\dotfill & $3.35_{-0.65}^{+0.76}$\\
                  ~~~$\log(g_{P})$\dotfill &Surface gravity\dotfill & $3.146_{-0.069}^{+0.065}$\\
      
\sidehead{RV Parameters:}
                                     ~~~$K$\dotfill &RV semi-amplitude (m/s)\dotfill & $2.90\pm0.25$\\
                 ~~~$M_P\sin i$\dotfill &Minimum mass ($M_{\oplus}$)\dotfill & $7.86\pm0.73$\\
           
\sidehead{Primary Transit Parameters:}
                ~~~$T_C$\dotfill &Time of mid-transit (\bjdtdb)\dotfill & $2456361.8050_{-0.0033}^{+0.0030}$\\
~~~$R_{P}/R_{*}$\dotfill &Radius of planet in stellar radii\dotfill & $0.0306\pm0.0014$\\
     ~~~$a/R_{*}$\dotfill &Semi-major axis in stellar radii\dotfill & $24.36_{-1.1}^{+0.97}$\\
                      ~~~$i$\dotfill &Inclination (degrees)\dotfill & $89.45_{-0.42}^{+0.37}$\\
                           ~~~$b$\dotfill &Impact Parameter\dotfill & $0.23_{-0.16}^{+0.18}$\\
                         ~~~$\delta$\dotfill &Transit depth\dotfill & $0.000934_{-0.000084}^{+0.000090}$\\
          ~~~$\tau$\dotfill &Ingress/egress duration (days)\dotfill & $0.00391_{-0.00030}^{+0.00054}$\\
                 ~~~$T_{14}$\dotfill &Total duration (days)\dotfill & $0.1238_{-0.0053}^{+0.0052}$

\enddata
\tablenotetext{-}{The parameters in italics (except the V magnitude) were used as input for the EXOFAST algorithm. The remaining parameters were obtained from the output of the EXOFAST fit.}
\label{tab:HD97658}
\end{deluxetable}


\begin{deluxetable}{lcc}
\tabletypesize{\footnotesize}
\tablecaption{Radial Velocities for HD\,97658
\label{tab:rvs}}
\tablewidth{0pt}
\tablehead{
\colhead{}         & \colhead{Radial Velocity}     & \colhead{Uncertainty}   \\
\colhead{BJD}   & \colhead{(m\,s$^{-1}$)}  & \colhead{(m\,s$^{-1}$)}
}
\startdata
 2453398.041747 &    \phantom{-}7.40 &    0.65               \\ 
 2453748.036160 &    \phantom{-}4.76 &    0.71                      \\ 
 2453806.962215 &    \phantom{-}2.51 &    0.71                \\ 
 2454085.159590 &   -4.83 &    0.79                \\ 
 2454246.878923 &   -2.33 &    0.72                      \\ 
 2454247.840558 &    -4.86 &     0.94  \\ 
 2454248.945454 &    -2.82 &     1.08  \\ 
 2454249.803197 &     0.22 &     1.07  \\ 
 2454250.840581 &     1.56 &     0.94  \\ 
 2454251.895304 &    -0.07 &     0.96  \\ 
 2454255.872627 &    -0.87 &     0.72  \\ 
 2454277.818152 &    -0.91 &     0.98  \\ 
 2454278.839136 &    -0.02 &     0.97  \\ 
 2454279.830756 &     2.28 &     0.99  \\ 
 2454294.764264 &    -5.63 &     1.10  \\ 
 2454300.742505 &    -0.60 &     1.02  \\ 
 2454304.762991 &    -5.95 &     1.17  \\ 
 2454305.759223 &    -5.06 &     0.80  \\ 
 2454306.772505 &    -4.31 &     0.97  \\ 
 2454307.747998 &    -0.93 &     0.77  \\ 
 2454308.749850 &     2.39 &     0.75  \\ 
 2454309.748488 &     1.32 &     1.04  \\ 
 2454310.744183 &    -0.21 &     1.03  \\ 
 2454311.744669 &     4.26 &     1.12  \\ 
 2454312.743176 &    -2.24 &     1.04  \\ 
 2454313.744948 &    -1.35 &     1.19  \\ 
 2454314.751499 &    -0.07 &     1.14  \\ 
 2454455.155080 &    -5.29 &     1.08  \\ 
 2454635.798361 &     0.75 &     0.96  \\ 
 2454780.126213 &    -3.91 &     1.10  \\ 
 2454807.091282 &    -7.71 &     1.20  \\ 
 2454808.158585 &    -4.91 &     1.25  \\ 
 2454809.144268 &    -1.59 &     1.14  \\ 
 2454810.025842 &     1.52 &     1.23  \\ 
 2454811.115461 &    -2.72 &     1.30  \\ 
 2454847.118970 &    -2.42 &     1.34  \\ 
 2454927.899109 &    -1.44 &     1.19  \\ 
 2454928.963980 &    -9.31 &     1.12  \\ 
 2454929.842494 &    -7.99 &     1.26  \\ 
 2454934.959381 &    -5.61 &     1.63  \\ 
 2454954.970889 &     0.00 &     1.08  \\ 
 2454955.924351 &     0.73 &     0.58  \\ 
 2454956.906013 &     1.26 &     0.58  \\
 2454963.966588 &     1.83 &     0.61  \\ 
 2454983.873218 &    -1.21 &     0.64  \\ 
 2454984.903863 &    -0.58 &     0.65  \\ 
 2454985.846210 &    -4.03 &     0.64  \\ 
 2454986.888827 &    -3.76 &     0.64  \\ 
 2454987.896299 &    -4.52 &     0.63  \\ 
 2454988.844192 &    -5.88 &     0.64  \\ 
 2455041.753229 &     0.80 &     1.32  \\ 
 2455164.116578 &     0.93 &     1.20  \\ 
 2455188.160043 &    -3.03 &     0.73  \\ 
 2455190.133815 &    -6.54 &     0.64  \\ 
 2455191.162113 &    -4.92 &     0.68  \\ 
 2455192.130000 &    -1.20 &     0.63  \\ 
 2455193.117044 &     0.90 &     0.65  \\ 
 2455197.145482 &    -2.04 &     0.65  \\ 
 2455198.064452 &    -3.58 &     0.67  \\ 
 2455199.090244 &    -3.69 &     0.65  \\ 
 2455256.958605 &     2.93 &     0.66  \\ 
 2455285.941646 &    -2.31 &     0.69  \\ 
 2455289.831867 &    -1.33 &     0.61  \\ 
 2455311.785608 &    -5.80 &     0.67  \\ 
 2455312.860423 &    -4.16 &     0.56  \\ 
 2455313.768309 &     0.20 &     0.64  \\ 
 2455314.782223 &     0.82 &     0.66  \\ 
 2455317.962910 &    -0.08 &     0.64  \\ 
 2455318.944742 &    -3.03 &     0.65  \\ 
 2455319.903509 &    -4.99 &     0.58  \\ 
 2455320.861260 &    -4.83 &     0.57  \\ 
 2455321.834736 &    -1.17 &     0.58  \\ 
 2455342.878772 &    -2.10 &     0.61  \\ 
 2455343.830638 &    -1.23 &     0.63  \\ 
 2455344.880812 &     1.12 &     0.65  \\ 
 2455350.781814 &    -4.60 &     0.56  \\ 
 2455351.884649 &     0.72 &     0.59  \\ 
 2455372.756478 &     2.28 &     0.58  \\ 
 2455373.784809 &    -0.70 &     0.56  \\ 
 2455374.759589 &     0.04 &     0.57  \\ 
 2455375.776842 &    -0.64 &     0.59  \\ 
 2455376.744926 &    -2.63 &     0.56  \\ 
 2455377.741425 &    -0.96 &     0.55  \\ 
 2455378.743929 &     2.85 &     0.61  \\ 
 2455379.791225 &     0.98 &     0.62  \\ 
 2455380.744606 &     5.58 &     0.59  \\ 
 2455400.743251 &    -0.31 &     0.66  \\ 
 2455401.770164 &    -1.35 &     1.38  \\ 
 2455403.738397 &    -3.92 &     0.70  \\ 
 2455404.737600 &    -4.22 &     0.64  \\ 
 2455405.740598 &    -5.58 &     0.63  \\ 
 2455406.738423 &    -3.22 &     0.60  \\ 
 2455407.758497 &     0.37 &     0.75  \\ 
 2455410.738763 &     5.05 &     0.64  \\ 
 2455411.734694 &     2.03 &     0.65  \\ 
 2455412.732758 &     2.06 &     1.15  \\ 
 2455413.736205 &     1.52 &     0.71  \\ 
 2455501.151034 &    -2.46 &     0.64  \\ 
 2455522.132828 &     1.48 &     0.63  \\ 
 2455529.170850 &   -12.62 &     1.53  \\ 
 2455543.149154 &     1.28 &     0.64  \\ 
 2455546.125032 &    -2.30 &     0.65  \\ 
 2455556.136099 &    12.63 &     0.70  \\ 
 2455557.076967 &    -3.11 &     0.67  \\ 
 2455559.127467 &    -2.44 &     0.77  \\ 
 2455585.099940 &    -3.61 &     0.64  \\ 
 2455605.987844 &    -5.75 &     0.66  \\ 
 2455606.982970 &    -3.25 &     0.65  \\ 
 2455607.982479 &    -2.00 &     0.74  \\ 
 2455614.039397 &    -0.19 &     0.64  \\ 
 2455614.875472 &     0.94 &     0.78  \\ 
 2455633.993494 &    -4.99 &     0.64  \\ 
 2455635.051246 &    -1.85 &     0.68  \\ 
 2455635.997009 &    -1.25 &     0.62  \\ 
 2455636.758083 &    -1.68 &     0.59  \\ 
 2455663.885585 &    -5.65 &     0.68  \\ 
 2455667.968022 &    -0.67 &     0.60  \\ 
 2455668.935707 &     1.36 &     0.62  \\ 
 2455670.839673 &    -3.71 &     0.61  \\ 
 2455671.811836 &    -2.62 &     0.62  \\ 
 2455672.799190 &    -1.02 &     0.64  \\ 
 2455673.805866 &    -0.23 &     0.60  \\ 
 2455696.875840 &    -1.48 &     0.63  \\ 
 2455697.796128 &    -2.55 &     0.60  \\ 
 2455698.800020 &    -3.67 &     0.59  \\ 
 2455699.806142 &    -6.66 &     0.61  \\ 
 2455700.825166 &    -4.76 &     0.62  \\ 
 2455703.777981 &    -1.62 &     0.60  \\ 
 2455704.749841 &     0.40 &     0.58  \\ 
 2455705.750495 &    -2.57 &     0.58  \\ 
 2455706.809245 &    -2.11 &     0.59  \\ 
 2455707.799643 &    -2.35 &     0.63  \\ 
 2455723.768188 &     1.27 &     0.42  \\ 
 2455728.755558 &    -3.21 &     0.60  \\ 
 2455731.795592 &     1.39 &     0.87  \\ 
 2455733.760561 &     0.91 &     0.58  \\ 
 2455734.783228 &    -2.46 &     0.55  \\ 
 2455735.790021 &    -2.10 &     0.54  \\ 
 2455738.754067 &    -4.23 &     0.58  \\ 
 2455751.746336 &     3.76 &     0.65  \\ 
 2455752.741618 &     1.77 &     0.62  \\ 
 2455759.747774 &    -0.60 &     0.61  \\ 
 2455760.738878 &    -0.60 &     0.63  \\ 
 2455761.744163 &     1.19 &     0.59  \\ 
 2455762.753715 &    -0.61 &     0.68  \\ 
 2455763.748343 &    -2.38 &     0.64  \\ 
 2455768.736141 &     0.15 &     0.62  \\ 
 2455770.749400 &     3.34 &     0.70  \\ 
 2455871.127171 &    -4.76 &     0.70  \\ 
 2455878.144281 &    -4.84 &     0.64  \\ 
 2455879.098924 &    -7.54 &     0.69  \\ 
 2455880.152597 &    -9.09 &     0.67  \\ 
 2455882.154960 &    -3.74 &     0.62  \\ 
 2455902.045678 &     0.46 &     0.63  \\ 
 2455903.044030 &     4.20 &     0.66  \\ 
 2455904.134955 &     5.19 &     0.62  \\ 
 2455905.066792 &     3.94 &     0.56  \\ 
 2455929.146269 &    -5.11 &     0.62  \\ 
 2455932.049104 &    -1.17 &     0.62  \\ 
 2455945.099002 &    -0.97 &     0.76  \\ 
 2455961.023265 &     6.60 &     0.65  \\ 
 2455967.944966 &    -0.47 &     0.67  \\ 
 2455972.956960 &    -0.51 &     0.61  \\ 
 2455990.905528 &     5.94 &     0.71  \\ 
 2455991.904565 &     2.46 &     0.70  \\ 
 2455997.010626 &    -0.79 &     0.62  \\ 
 2455999.795911 &     3.83 &     0.67  \\ 
 2456018.905001 &     6.66 &     0.66  \\ 
 2456019.968956 &     2.26 &     0.62  \\ 
 2456027.799773 &     2.93 &     0.66  \\ 
 2456102.749354 &     4.11 &     0.57  \\ 
 2456111.744838 &     4.65 &     0.64  \\ 
 2456145.735429 &     0.73 &     0.75  \\ 
 2456266.103384 &     2.22 &     0.70  \\

\enddata
\end{deluxetable}

\end{document}